%% file: Fragmentation-of-shells.tex
\documentclass[%
aip,
amsmath,amssymb,
reprint,%
]{revtex4-1}
\usepackage{graphicx}
\usepackage{dcolumn}
\usepackage{bm}

\usepackage[utf8]{inputenc}
\usepackage[T1]{fontenc}
\usepackage{mathptmx}
\usepackage[colorlinks,linkcolor=black,anchorcolor=black,citecolor=blue,urlcolor=blue]{hyperref}
\draft 

\begin{document}


\title{Fragmentation of shells: An analogy with the crack formation in tree bark} 



\author{Chuang-Shi Shen}
\email[]{chshshen@nwpu.edu.cn}
\author{Chao Zhang}
\email[Corresponding Author]{: chaozhang@nwpu.edu.cn}
\affiliation{ School of Aeronautics, Northwestern Polytechnical University, Xi'an 710072, China.}
\affiliation{ Shaanxi Key Laboratory of Impact Dynamics and Its Engineering Applications, Northwestern Polytechnical University, Xi'an 710072, China.}
\affiliation{ Joint International Laboratory of Impact Dynamics and Its Engineering Applications, Northwestern Polytechnical University, Xi'an 710072, China.}
\author{Xiaosheng Gao}
\affiliation{ Department of Mechanical Engineering, The University of Akron, Akron 44325, United States.}
\author{Yulong Li}
\affiliation{ School of Aeronautics, Northwestern Polytechnical University, Xi'an 710072, China.}
\affiliation{ Shaanxi Key Laboratory of Impact Dynamics and Its Engineering Applications, Northwestern Polytechnical University, Xi'an 710072, China.}
\affiliation{ Joint International Laboratory of Impact Dynamics and Its Engineering Applications, Northwestern Polytechnical University, Xi'an 710072, China.}


\begin{abstract}
How does the shell explode into a series of fragments? The well-accepted explanation is Mott’s theory, which considers fragmentation of shells is a random process controlled by defects. However, Mott’s theory is inadequate in its energy-equivalence assumption and is incapable in explaining the saturation of fragment length with the increase of expansion velocity. In this letter, we present a theory to explain the physical mechanism of fragmentation of shells and propose a high-efficient model for predicting the number of necks after fragmentation. We recognize that the pressurized gas is in close contact with the shell layer during an explosion and the internal pressure is transferred to the shell layer following the shear-lag principle. The fragmentation problem of shells is then analogous to the cracking behavior of the tree bark, and the closed-form solutions are obtained to describe the relationship between the expansion velocity and the number of necks with consideration of the strain-rate dependent strength of the shell material. The theoretical results show excellent correlation with the experimental results.
\end{abstract}


\maketitle 

Fragmentation of shells is an classic and elementary problem in fracture dynamics, corresponding to the phenomena that a shell body subjected to external pressure forces resulting in the multiple fracturing of the body. The simplified one-dimensional ring is generally introduced to investigate the failure behavior of fragmentation of shells, as shown in Fig.1(a). The well-accepted mechanism behind this was proposed by Sir Nevill Mott  \cite{1943mott1,1943mott2,mott1947fragmentation} between 1943 to 1947. As shown in Fig. 1(b), he postulated that fracture begins at one of a number of “weak points” distributed throughout the material, and when instantaneous fracture occurs at one weak point, stress waves (Mott’s waves) propagate away from the fracture releasing the stress in the neighborhood. Within the region of the specimen subjected to the action of the released wave, the material does not continue to stretch and thus failure is precluded. The size of the fragments is then determined by the travelled distance of Mott’s wave. Since it is impossible to determine the travel time of Mott’s wave, Mott made a simple assumption that the surface energy required to generate new fracture surfaces is provided by the local kinetic energy of the fragments (This is known as Mott’s energy-equivalence assumption). Accordingly, a mean length formula of fragment is proposed based on the law of energy conservation,  $l = {\left( {{{24{G_c}{r^2}} \mathord{\left/{\vphantom {{24{G_c}{r^2}} {\rho {V^2}}}} \right.\kern-\nulldelimiterspace} {\rho {V^2}}}} \right)^{{1 \mathord{\left/{\vphantom {1 3}} \right.\kern-\nulldelimiterspace} 3}}}$, where ${{G_c}}$, $\rho $, $r$ and $V$ are the fracture energy, density, ring radius and expansion velocity, respectively. Grady and co-workers  \cite{Kipp1985Dynamic,Grady2003A,Dr2006Fragmentation} further popularized Mott's theory between the early 1980s to the late 2010s. A major defect of Mott’s theory is the energy-equivalence assumption, as it is in contrast with the energy-conversion principle, which suggests that the work done by the explosive loads should equal to the sum of the energy required to generate new fracture surfaces, the kinetic energy of the fragments and other dissipated energy. Another important factor is the strain rate effect, which is not considered in Mott’s theory. Zhang and Ravi-Chandar \cite{zhang2006dynamics,zhang2008dynamics} revealed that the onset of fragmentation is caused by the development of a part of the nucleated necks, thus the number of necks formed exceeds the number of fragments, and the variance of the fragment length decreases with the increase of the expanding velocity suggesting its value becomes zero when the expanding velocity is sufficiently high. More recently, both experiments \cite{kahana2015microstructural,zhang2017,zhang2018experimental,ma2014dynamic} and finite element simulations \cite{rodriguez2013identification,n2018random} have shown that defects barely affect the average neck spacing and the neck spacing no long reduces at very high expanding velocity, which could not be explained by the statistical fragmentation theory of Mott. Thus it has raised the question that the localization process become deterministic at high expansion velocities. So far, the physical mechanism of fragmentation of shells remains unclear. In this letter, we present a novel theory to explain the physical mechanism of fragmentation of shells and propose an efficient theoretical model for predicting the number of necks after fragmentation.
\begin{figure}[b]
	\includegraphics[scale=1]{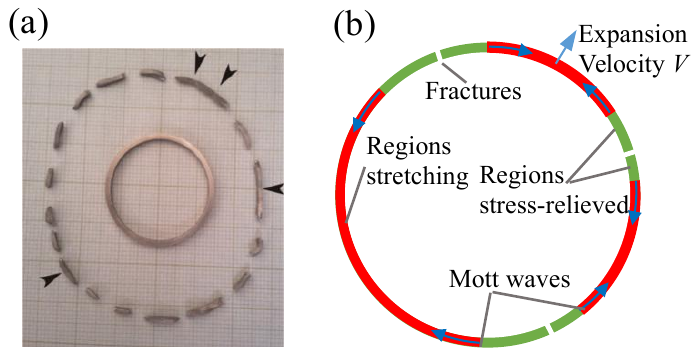}
	\caption{\label{fig:epsart} (a) Example of a ring specimen before test and the fractured segments after explosion; several fragments contain necks that did not develop to fracture (indicated by arrows) \cite{kahana2015microstructural}. (b) Scheme for the statistical fragmentation theory developed by Mott.}
\end{figure}

\begin{figure*}
	\includegraphics[scale=1]{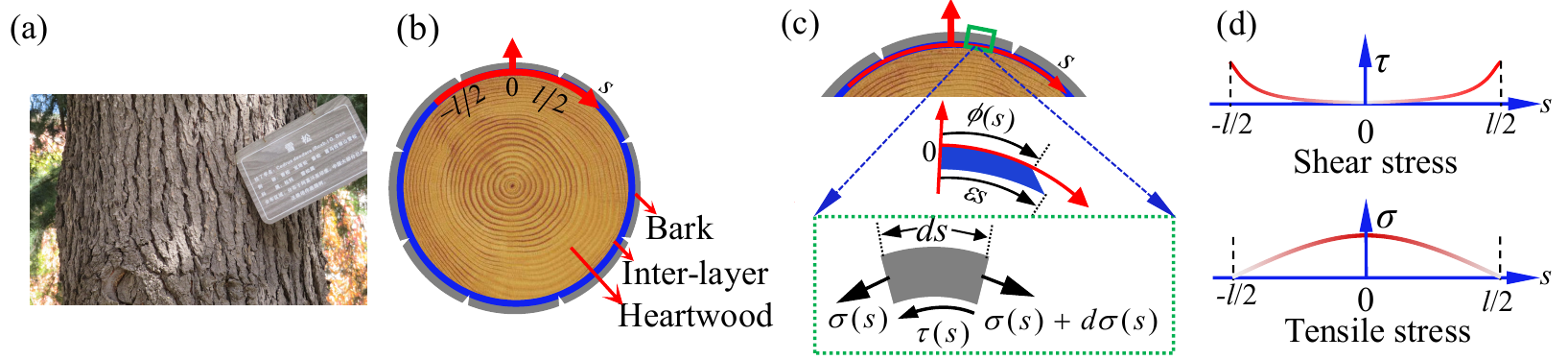}
	\caption{\label{fig:wide} The mechanical mechanism of crack formation for the tree bark. (a) The bark of a cedar tree, where the cracks in the bark are almost equally spaced along the hoop direction. (b) Scheme of the tree cross section, which contains three parts: the bark, the heartwood, and the inter-layer between the bark and the heartwood. (c) The free-body diagram of a bark segment. (d) Stress distribution in the inter-layer and the bark, where the tensile stress is the greatest in the middle of each bark segment. }
\end{figure*}

The principles governing the fragmentation process can be summarized as the following three criteria: 1) crack initiates in structure where stress exceeds the strength of the material, 2) stress is released as a result of fracture so that it falls below the strength of the material in other parts of the structure, and 3) energy consumed by forming the cracks is minimized. The second criterion promotes more cracks being generated while the third criterion suggests the opposite. The compromise between the two competing criteria determines the number and size of fragments.

Before studying the fragmentation of shells, let’s first introduce a nature phenomenon: the cracking behavior of the tree bark as show in Fig. 2(a). The mechanism of crack formation in the tree bark is illustrated schematically in Fig. 2(b-d). Since the growth rate of the heartwood is higher than that of the tree bark, an expansion strain develops in the heartwood as the tree grows. This expansion strain is then transferred to the bark through the interfacial shear stress, based on the shear-lag principle \cite{Cox1951The,Mcguigan2003An}. For a single crack island generated during the tree growth period, from center to edge, the interfacial shear stress increases from zero to a peak value while the normal tensile stress decreases from a peak value to zero due to the free-edge effect\cite{1970}, as shown in Fig. 2(d). When the tensile stress in the middle point of bark segment is slightly lower than the tensile strength, the length of this segment will satisfy the requirements of both the second and third fragmentation criterion at the same time.

Similar fracture mode is also commonly seen in other layered materials \cite{adda2001fracture,bai2000explanation,Faurie2017Fragmentation,shabahang2016controlled} and layer-like structures, such as fiber reinforced composites \cite{berthelot2003transverse}, painted structures \cite{giorgiutti2016painting}, craquelure in ceramics \cite{bohn2005four}, dry bed-like fracture of lithium-ion battery \cite{beaulieu2001colossal}, thermal shock crack \cite{song2010enhanced}, shrinkage cracks developed during cooling  \cite{Aydin1988Evoluton,Kumar2017Stress},  and water droplets impacting a cold surface \cite{ghabache2016frozen}.

\begin{figure*}
	\includegraphics[scale=1]{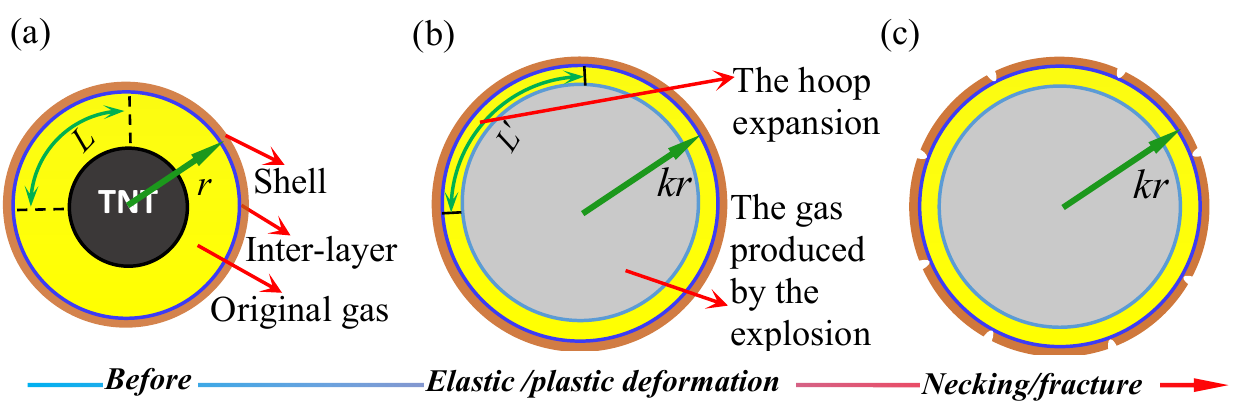}
	\caption{\label{fig:wide}  Fragmentation of shells. (a) Scheme of the shell before explosion, where the original radius of shell equals to $r$  and an inter-layer is assumed to present between the gas and shell. (b) Scheme for the instant of explosion when the shell undergoes elastic/plastic deformation but the necking/fracture has not occurred yet. The internal gas is significantly pressurized, resulting in a hoop expansion. The radius of the shell increases from  $r$ to $kr$ . (c) Scheme for the instant when necking/fracture just occurred but gas has not leaked yet. The shell and high pressure gas are still in close contact, while the instantaneous tensile stress in the middle of each segment is equal to the tensile strength of shell material, and the tensile stress in the two ends of each segment is zero.}
\end{figure*}

In this work, fragmentation of shells is considered to be analogous to the crack formation process in layered structures, for example, the tree bark. The shell is equivalent to the bark of the tree, and the high pressure gas produced by the explosion is equivalent to the growing trunk. The high-pressure gas and the shell stick tightly with each other before fragmentation, which is equal to the bark and heartwood adhere to each other during the growth of the tree. The scheme for the fragmentation process of shells is shown in Fig. 3(a-c). At the instant of explosion, the internal gas is significantly pressurized, resulting in a hoop expansion as shown in Fig. 3(b), which is transferred to the shell through interfacial shear stress, similar to the transferring of stress from heartwood to tree bark. Thus the fragmentation problem can be analogy with the cracking problem of a multi-layer structure, and can be solved using the approach reported by Mcguigan and coworkers \cite{Mcguigan2003An}. The only difference between the fragmentation of shells and bark cracking is that one happens rapidly while the other happens in a much slower manner.

Consider a shell as shown in Fig. 3(a) with a radius $r$, thickness $h$ and cross-sectional area $A$. Let $n$ be the number of fragments after fragmentation, $\rho $ be the density, ${G_c}$ be the energy per unit area required to form a crack, $E$ be the elastic modulus and $V$ be the expansion velocity. Based on conservation of energy, the work done by the explosion, $W$, is equal to the sum of the surface energy required to generate new fracture surfaces, $\Gamma $, the kinetic energy of the fragments, $T$, and the dissipated energy due to plastic deformation, ${E_p}$,  i.e., $W = T + \Gamma  + {E_p}$, where $T = \pi rA\rho {V^2}$ and $\Gamma  = 2n{G_c}A$. For high-velocity explosion problem, $T \gg \Gamma  + {E_p}$, and therefore,  almost all of the work done by the explosion is converted into the kinetic energy , i.e., $W \approx T$. Thus the fragmentation of plastic shell with radius of $r$ is approximately equal to a brittle shell with radius $kr$ under the action of explosive loads and then fracture into many fragments.

Let ${p_{\max }}$ be the maximum impulse pressure acting on the inner side of the shell and ${t_d}$ be the duration of the explosion. During an explosion, the impulse pressure ${p}$ acting on the shells is very small for the initial stage (0 to ${t_e}$),  followed by an approximately linear increase (${t_e}$ to ${t_d}$), and then drop suddenly to zero \cite{kahana2015microstructural,zhang2017}. The explosion period can be approximate as ${t_d} - {t_e}$. Thus, we have $p \approx \left( {{{{p_{\max }}} \mathord{\left/{\vphantom {{{p_{\max }}} {\left( {{t_d} - {t_e}} \right)}}} \right.\kern-\nulldelimiterspace} {\left( {{t_d} - {t_e}} \right)}}} \right)\left( {t - {t_e}} \right)$. The impulse-momentum theorem suggests that $2\pi r\int_{{t_e}}^{{t_d}} {pdt}  \approx 2\pi rh\rho V$, which results in ${p_{\max }} \approx {{2\rho Vh} \mathord{\left/
		{\vphantom {{2\rho Vh} {({t_d} - {t_e})}}} \right.
		\kern-\nulldelimiterspace} {({t_d} - {t_e})}}$. Hence, the maximum average hoop strain in the shell during the explosion is $\overline \varepsilon   \approx {{2\rho rV} \mathord{\left/
		{\vphantom {{2\rho rV} {E({t_d} - {t_e})}}} \right.
		\kern-\nulldelimiterspace} {E({t_d} - {t_e})}}$. Considering the tight contact between the high-pressure gas and the shell, the hoop expansion strain in the outermost layer of the high-pressure gas is considered to be uniform along the hoop direction with a value of $\varepsilon  \approx {{2\rho rV} \mathord{\left/
		{\vphantom {{2\rho rV} {E({t_d} - {t_e})}}} \right.
		\kern-\nulldelimiterspace} {E({t_d} - {t_e})}}$.

The tensile stress in the thin shell is assumed to be uniform along the thickness direction. Based on the free-body diagram shown in Fig. 2(c), the basic equation relating the inter-facial shear stress $\tau (s)$ to the tensile stress $\sigma (s)$ in the shell is related by the following equation
\begin{equation}h\frac{{d\sigma (s)}}{{ds}} =  - \tau (s)\end{equation}

The constitutive equation of the inter-layer is
\begin{equation}\tau (s) = {{\overline G (\varepsilon s - \phi (s))} \mathord{\left/
		{\vphantom {{\overline G (\varepsilon s - \phi (s))} {\overline h }}} \right.
		\kern-\nulldelimiterspace} {\overline h }}\end{equation}
where $\phi (s)$ is the elastic displacement of shell, $\overline G $ is shear modulus of inter-layer, $\overline h $ is the thickness of inter-layer. The tensile stress in the fragment, $\sigma (s)$,  is related to the elastic displacement of shell through
\begin{equation}\sigma (s) = E\frac{{d\phi (s)}}{{ds}}\end{equation}

Combining Eqs. (1), (2) and (3) yields a differential equation below
\begin{equation}\frac{{{d^2}\phi (s)}}{{d{s^2}}} - \frac{{\overline G }}{{Eh\overline h }}\phi (s) + \frac{{\overline G }}{{Eh\overline h }}\varepsilon s = 0\end{equation}

The general solution of Eq. (4) is
\begin{equation}
\phi (s) = C\sinh (\sqrt {{{\overline G } \mathord{\left/
			{\vphantom {{\overline G } {Eh\bar h}}} \right.
			\kern-\nulldelimiterspace} {Eh\bar h}}} s) + \varepsilon s
\end{equation}
where $C$ is to be determined from the boundary conditions. Substituting Eq. (5) to Eq. (3), the tensile stress of the shell can be expressed as
\begin{equation}
\sigma (s) = E\left[ {\left( {C\sqrt {{{\overline G } \mathord{\left/
					{\vphantom {{\overline G } {Eh\bar h}}} \right.
					\kern-\nulldelimiterspace} {Eh\bar h}}} } \right)\cosh \left( {\sqrt {{{\overline G } \mathord{\left/
					{\vphantom {{\overline G } {Eh\bar h}}} \right.
					\kern-\nulldelimiterspace} {Eh\bar h}}} s} \right) + \varepsilon } \right]
\end{equation}

Considering the instant when necking has just occurred but gas has not leaked yet as shown in Fig. 3(c), the coefficient $C$ in Eq. (6) can be determined by the boundary condition $\sigma ({l \mathord{\left/{\vphantom {l 2}} \right.\kern-\nulldelimiterspace} 2}) \approx 0$

\begin{equation}
C \approx \frac{{ - \varepsilon }}{{\sqrt {{{\overline G } \mathord{\left/
					{\vphantom {{\overline G } {Eh\bar h}}} \right.
					\kern-\nulldelimiterspace} {Eh\bar h}}} \cosh \left( {\sqrt {{{\overline G } \mathord{\left/
						{\vphantom {{\overline G } {Eh\bar h}}} \right.
						\kern-\nulldelimiterspace} {Eh\bar h}}} {l \mathord{\left/
					{\vphantom {l 2}} \right.
					\kern-\nulldelimiterspace} 2}} \right)}}
\end{equation}

The maximum tensile stress occurs in the middle of segment as shown in Fig. 2(d). We look for a solution where this maximum stress is equal to the tensile strength, which corresponds to a segment length satisfying both requirements of the second and third criterion for fragmentation. This can be written as $\sigma (0) = {\sigma _{\max }}(s) = {\sigma ^ * }(\dot \varepsilon )$, where ${\sigma ^ * }(\dot \varepsilon )$ is the strain-rate dependent tensile strength and $\dot \varepsilon  = {V \mathord{\left/{\vphantom {V r}} \right.\kern-\nulldelimiterspace} r}$. This results in the following equation
\begin{equation}
\left[ {\frac{{ - \varepsilon }}{{\cosh \left( {\sqrt {{{\overline G } \mathord{\left/
							{\vphantom {{\overline G } {Eh\bar h}}} \right.
							\kern-\nulldelimiterspace} {Eh\bar h}}} {{\mathop l\nolimits^ *  } \mathord{\left/
						{\vphantom {{\mathop l\nolimits^ *  } 2}} \right.
						\kern-\nulldelimiterspace} 2}} \right)}}\cosh (0) + \varepsilon } \right] = {\sigma ^ * }(\dot \varepsilon )
\end{equation}

Solving Eq. (8), the length of the neck segment can be expressed as follows
\begin{equation}{l^ * } \approx 2\sqrt {\frac{{Eh\overline h }}{{\overline G }}} {\cosh ^{ - 1}}\left[ {\frac{1}{{1 - {{{\sigma ^ * }(\dot \varepsilon )} \mathord{\left/
					{\vphantom {{{\sigma ^ * }(\dot \varepsilon )} {E\varepsilon }}} \right.
					\kern-\nulldelimiterspace} {E\varepsilon }}}}} \right]\end{equation}

Consequently, the number of necks after fragmentation can be calculated by
\begin{equation}N \approx \pi rk\sqrt {\frac{{\overline G }}{{\overline h }}\frac{1}{{Eh}}} {\left[ {{{\cosh }^{ - 1}}(\frac{1}{{1 - {{{\sigma ^ * }(\dot \varepsilon )({t_d} - {t_e})} \mathord{\left/{\vphantom {{{\sigma ^ * }(\dot \varepsilon )({t_d} - {t_e})} {(2\rho rV)}}} \right.\kern-\nulldelimiterspace} {(2\rho rV)}}}})} \right]^{ - 1}}
\end{equation}

\begin{figure}[b]
	\includegraphics[scale=1]{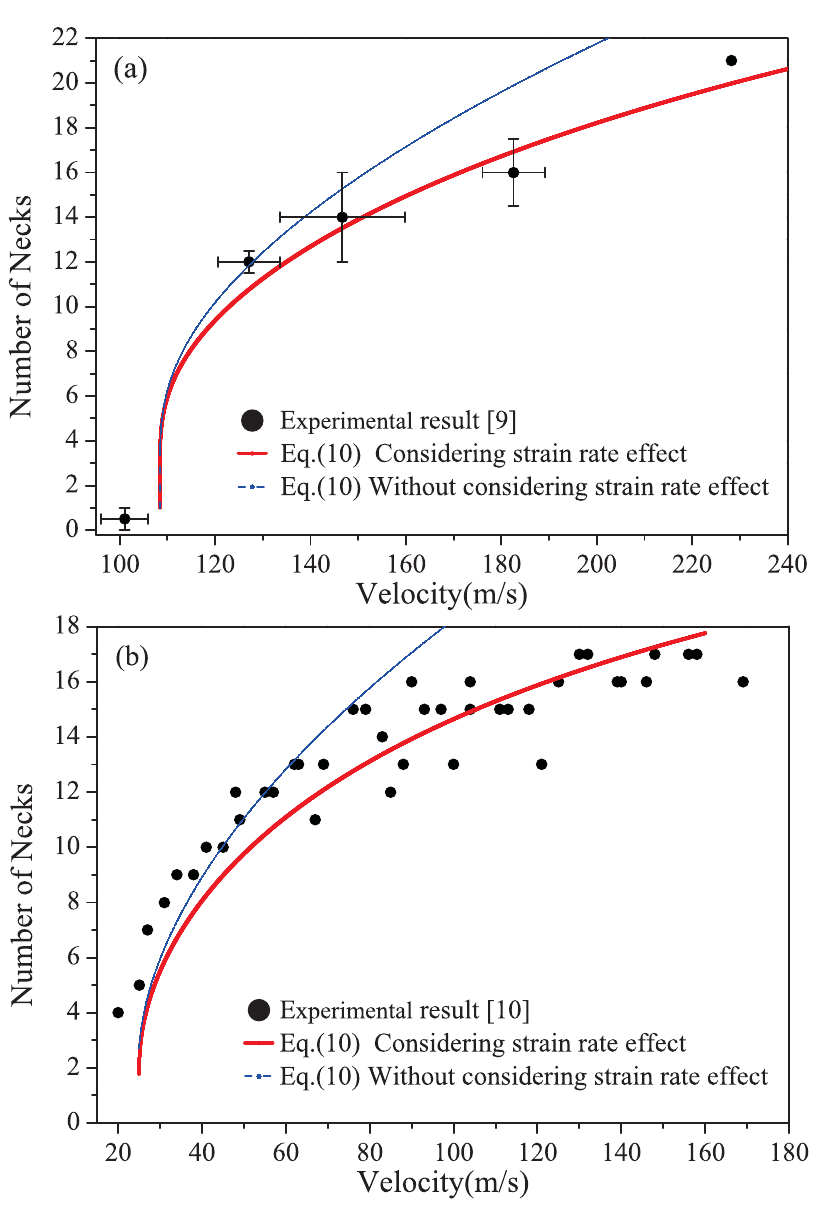}
	\caption{\label{fig:epsart} Relationship for number of necks against expansion velocity. (a) AZ31 magnesium alloy, (b) 1060 pure aluminum. The increasing rate for the number of necks slows down gradually with the increasing of expansion velocity. The strain-rate effect has a great influence on the number of necks, especially for high expansion velocities.}
\end{figure}

There are three important parameters in Eq. (10) that need to be determined, the explosion period ${t_d} - {t_e}$, the rate-dependent strength ${\sigma ^ * }(\dot \varepsilon )$ and ${{{k^2}\overline G } \mathord{\left/{\vphantom {{{k^2}\overline G } {\overline h }}} \right.\kern-\nulldelimiterspace} {\overline h }}$ . Kahana et al \cite{kahana2015microstructural} and Zhang et al. \cite{zhang2017,zhang2018experimental} conducted ring expansion tests of AZ31 magnesium alloy rings ($r = 16.3{\rm{mm}}$, $h = 1.7{\rm{mm}}$, $E = 45{\rm{GPa}}$, $v = 0.35$, ${\sigma ^ * } = 240{\rm{MPa}}$, $\rho  = 1.8{{{\rm{Kg}}} \mathord{\left/{\vphantom {{{\rm{Kg}}} {{{\rm{m}}^{\rm{3}}}}}}\right.\kern-\nulldelimiterspace} {{{\rm{m}}^{\rm{3}}}}}$) and 1060 aluminum rings ($r = 16{\rm{mm}}$, $h = 1.5{\rm{mm}}$, $E = 72{\rm{GPa}}$, $v = 0.33$, ${\sigma ^ * } = 90{\rm{MPa}}$, $\rho  = 2.68{{{\rm{Kg}}} \mathord{\left/{\vphantom {{{\rm{Kg}}} {{{\rm{m}}^{\rm{3}}}}}} \right.\kern-\nulldelimiterspace} {{{\rm{m}}^{\rm{3}}}}}$) respectively. For the test of AZ31 magnesium alloy rings \cite{kahana2015microstructural}, the load increased from zero to a certain value over a very short period of time and the ring finally segmented in about $20\mu s$.   For the test of 1060 pure aluminum rings \cite{zhang2017}, the pressure acting on the shells was very small in the initial stage, followed by a linear increase $50{\rm{\mu s}}$ to $70\mu s$, and then dropped suddenly to zero. Therefore, for both experiments\cite{kahana2015microstructural,zhang2017}, we assume the  explosion period to be approximately $20\mu s$.

Shell fragmentation is a typical dynamic fracture problem, and the strain rate effect must be considered. Yu et al \cite{shui2013strain} found that the tensile strengths of many metallic materials increase approximately linearly with the increase of strain rate over the strain rate sensitive range. This linear relationship is assumed applicable to the materials considered in this study. Based on the dynamic mechanical tests by Feng et al \cite{feng2014experimental}, Xu et al \cite{xu2017tensile} and Khan and Huang \cite{khan1992experimental}, the tensile strength of AZ31 at strain rate of ${10^4}{{\rm{s}}^{{\rm{ - 1}}}}$ is $ \sim 1.7$ times of its quasi-static value \cite{feng2014experimental,xu2017tensile} and the tensile strength of 1060 aluminum at strain rate of $6000{{\rm{s}}^{{\rm{ - 1}}}}$ is $ \sim 2$ times of its quasi-static value \cite{khan1992experimental}.

Now that all parameters in Eq. (10) have been determined, except for ${{{k^2}\overline G } \mathord{\left/{\vphantom {{{k^2}\overline G } {\overline h }}} \right.\kern-\nulldelimiterspace} {\overline h }}$, which can be calibrated by matching the model prediction with experimental data. The calibration results suggest that ${{{k^2}\overline G } \mathord{\left/{\vphantom {{{k^2}\overline G } {\overline h }}} \right.\kern-\nulldelimiterspace} {\overline h }}{\rm{ = 27}}{\rm{.78}} \times {\rm{1}}{{\rm{0}}^{\rm{3}}}{{{\rm{GPa}}} \mathord{\left/{\vphantom {{{\rm{GPa}}} {\rm{m}}}} \right.\kern-\nulldelimiterspace} {\rm{m}}}$ for AZ31 magnesium alloy rings, ${{{k^2}\overline G } \mathord{\left/{\vphantom {{{k^2}\overline G } {\overline h }}} \right.\kern-\nulldelimiterspace} {\overline h }}{\rm{ = 8}}{\rm{.89}} \times {\rm{1}}{{\rm{0}}^{\rm{3}}}{{{\rm{GPa}}} \mathord{\left/{\vphantom {{{\rm{GPa}}} {\rm{m}}}} \right.\kern-\nulldelimiterspace} {\rm{m}}}$ for 1060 aluminum rings.

Fig. 4 compares the predicted number of necks with the experimental observations by Kahana et al \cite{kahana2015microstructural} and Zhang et al \cite{zhang2017}. The predictions by Eq. (10) show good agreements with experimental results \cite{kahana2015microstructural,zhang2017}. The number of necks increases dramatically at low expansion velocities, but the increasing rate slows down gradually as the expansion velocity further increases. It is expected that the number of necks will reach a saturation value when the expansion velocity goes beyond a certain threshold. The flatten tendency of fragmentation can be explained in two aspects. Firstly, as form Eq. (6), the expansion strain $\varepsilon $ (the expansion velocity $V$) has a nonlinear exponential relationship with the segment length for a constant value of $\sigma \left( s \right)$. Secondly, the increase in expansion velocity corresponds to an increase of tensile strength of material. As shown in Fig. 4, with the strain rate effect included, the predicted number of necks is significantly smaller than the predicted number without considering strain rate effect.

In summary, this letter presents a new theoretical model to explain the physical mechanism of fragmentation of shells. At the instant of explosion, a two-layer structure is formed, where the outer layer is the shell and the high pressure gas generated by the explosion can be considered as the internal layer. Thus, the fragmentation process of shells becomes similar to the process of crack formation for the tree bark. A theoretical model for predicting the number of necks after fragmentation is proposed based on the shear-lag principle, with consideration of the strain-rate dependent strength properties. The model is calibrated for the 1060 pure aluminum and AZ31 magnesium alloy, and the predictions show good agreement with the experimental results. Furthermore, it is shown that the strain rate effect strongly influences fragmentation of shells.

The physical mechanism presented in this letter can also be used to explain the other fragmentation, such as a bullet impact on windows \cite{vandenberghe2013star,aastrom1997fracture,verberck2013materials,Rebeca2014Fragmentation}, thin layers of suspensions of non-Brownian particles that experience an impact \cite{roche2013dynamic}, popping balloons \cite{moulinet2015popping},  the failure of tubular energy absorber \cite{Hou2015Pressurised}, etc. Due to the space-time effect, the hoop stress near the impact region is significantly higher than the hoop stress far away from the impact region at the moment of impact, and a layer-like structure is formed. Consequently, the physical mechanism as show in Fig. 1 take charge of the rest of the process, and the star-shaped crack pattern is formed eventually.\\

C.S., C.Z. and Y.L. acknowledge National Natural Science Foundation of China (Grant No. 51706187 and 11772267) for financial support. C.S. would like to show deepest gratitude to his PhD supervisor, Dr. Xiaoping Han, for his enlightening instruction and suggestions. C.S. thank Wenjun Wang and Ruqian Guo for helpful discussions.\\
\\

\noindent$\displaystyle{\textbf{REFERENCES}}$
\input{Fragmentation-of-shells.bbl}
\end{document}

%% file: Fragmentation-of-shells.bbl
%